\documentclass[a4paper]{revtex4}
\usepackage{fancyhdr}
\usepackage{amsmath}
\usepackage{amsmath,amssymb,ifpdf,slashed,multirow,feynmp}
\ifpdf                                     
  \usepackage{graphicx}                    
\else                                      
\fi

\newcommand{\TeV}{\,{\rm TeV}}
\newcommand{\GeV}{\,{\rm GeV}}

\newcommand{\invfb}{\,{\rm fb^{-1}}}

\newcommand{\vev}[1]{\left\langle #1\right\rangle}
\newcommand{\s}[1]{_{\rm #1}}

\newcommand{\Order}{\mathop{\mathcal{O}}}

\newcommand{\gmt}{(g-2)_\mu}
\def\sgn{\mathop{\mathrm{sgn}}}

\makeatletter
\def\EE{\@ifnextchar-{\@@EE}{\@EE}}
\def\@EE#1{\ifnum#1=1 \times\!10 \else \times\!10^{#1}\fi}
\def\@@EE#1#2{\times\!10^{-#2}}
\makeatother

\begin{document}

\title{LHC SUSY searches after the Higgs discovery: respecting the muon $\boldsymbol{g-2}$}
\author{Sho Iwamoto%
  \footnote{Research Fellow of the Japan Society for the Promotion of Science.}%
  \footnote{This report is based on Ref.~\cite{Endo:2013bba}, which was completed in collaboration with Dr.~Endo, Prof.~Hamaguchi, and Mr.~Yoshinaga.}%
  \footnote{The presentation file of this talk is available at \url{http://en.misho-web.com/phys/talks.html}. }%
}
\affiliation{Department of Physics, The University of Tokyo, Tokyo 113--0033, JAPAN%
  \footnote{As of February 2013. In April 2013, SI moved to {\em Kavli IPMU, The University of Tokyo, Chiba 277--8583, JAPAN}.}%
}
\begin{abstract}
SUSY searches at the LHC as well as the $126\GeV$ Higgs boson indicate that superparticles, especially squarks and gluinos, are not so light as we expected.
It is important to investigate SUSY searches which do not rely on the colored superparticles.

As a clue for the investigation, we focus on the muon $g-2$ anomaly, which can be explained by the SUSY contributions if some of neutralinos, charginos, and sleptons are as light as $\Order(100)\GeV$.
We propose the $\gmt$-motivated MSSM as a benchmark model, where squarks are decoupled but the superparticles corresponding to the muon $g-2$ are light enough to explain the anomaly.
We also interpret the up-to-date results of LHC SUSY searches, and obtain experimental constraints on the model.

We show searches for direct production of charginos and neutralinos work very well against the scenario, but several regions are not only remain uncovered but even found challenging to be searched for at the LHC.
It is ascertained that, in order to draw out latent potential of the LHC, strategies to attack these regions should be developed.
\end{abstract}

\maketitle
\thispagestyle{fancy}


\begin{figure}[b]
 \begin{center}\vspace{\baselineskip}
 \begin{fmffile}{feyn_neut_mul_g-2}
\begin{fmfgraph*}(125,65)
\fmfleft{d1,p1,d2,d3,d4}\fmfright{d5,p2,d6,d7,gc}
\fmf{fermion,label=$\mu$,l.s=right}{p1,x1}
\fmf{fermion,label=$\mu$,l.s=right}{x2,p2}
\fmf{xgaugino,tension=0.5,lab=$\tilde\chi^0$,l.s=left}{x2,x1}
\fmf{phantom,tension=5}{gc,gb}
\fmf{phantom,tension=1}{gb,d1}
\fmfposition
\fmf{phantom,left,tag=1,lab=$\tilde\mu$,l.s=right}{x1,x2}
\fmfipath{p[]}
\fmfiset{p1}{vpath1(__x1,__x2)}
\fmfi{photon}{point length(p1)/3*2 of p1 -- vloc(__gb)}
\fmfv{label=$\gamma$}{gb}
\fmfi{dashes}{p1}
\end{fmfgraph*}
\end{fmffile}
\hspace{10pt}
 \begin{fmffile}{feyn_neut_nu_g-2}
\begin{fmfgraph*}(125,65)
\fmfleft{d1,p1,d2,d3,d4}\fmfright{d5,p2,d6,d7,gc}
\fmf{fermion,label=$\mu$,l.s=right}{p1,x1}
\fmf{fermion,label=$\mu$,l.s=right}{x2,p2}
\fmf{dashes,tension=0.5,lab=$\tilde\nu_\mu$,l.s=left}{x2,x1}
\fmf{phantom,tension=5}{gc,gb}
\fmf{phantom,tension=1}{gb,d1}
\fmfposition
\fmf{phantom,left,tag=1,lab=$\tilde\chi^-$,l.s=right}{x1,x2}
\fmfipath{p[]}
\fmfiset{p1}{vpath1(__x1,__x2)}
\fmfi{photon}{point length(p1)/3*2 of p1 -- vloc(__gb)}
\fmfv{label=$\gamma$}{gb}
\fmfi{xgaugino}{p1}
\end{fmfgraph*}
 \end{fmffile}
\caption[The MSSM dominant contributions to the muon $g-2$ (in the mass eigenstates).]{The diagrams of the MSSM dominant contributions to the muon $g-2$.}
\label{fig:MSSM_g-2}
\end{center}
\end{figure}
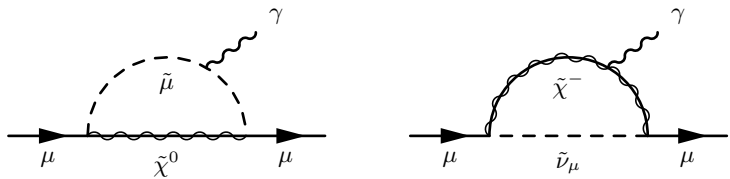

\section{Introduction}
Let us clarify the situation of the supersymmetry (SUSY), the most promising candidate for physics beyond the Standard model.
A Higgs boson was discovered at the LHC, and its mass was revealed to be approximately $126\GeV$~\cite{aad:2012tfa,Chatrchyan:2012ufa}.
This discovery completed the Standard Model, but at the same time, ascertained that the Standard Model suffers from the hierarchy problem~\cite{HierarchyProblem}.
The necessity of the SUSY as the solution to the hierarchy problem is definite.
However, no signatures of the SUSY have been captured at the LHC.
The squarks lighter than $\sim1.5\TeV$, and the gluinos lighter than $\sim950\GeV$ are now excluded by SUSY searches focusing on colored particle pair-production~\cite{ATLAS2012109,Chatrchyan:2012jx,Chatrchyan:2012mfa}.
Moreover, the mass of the Higgs boson, $126\GeV$, favors that the top-squarks as heavy as $\Order(1\text{--}10)\TeV$~\cite{MSSMLoopHiggsMass}.
All the results obtained at the LHC indicate that squarks are out of the LHC reach.

Here we focus on another insufficiency of the Standard Model, the anomaly on the anomalous magnetic moment of the muon, $\gmt$.
The measured value of $\gmt$ is deviated at $3\sigma$-level from the theoretical value based on the Standard Model:
\begin{equation}
\Delta a_\mu \equiv a_\mu\text{(exp)} - a_\mu\text{(SM)}=(26.1\pm8.0)\EE{-10},
\end{equation}
where $a_\mu\equiv (g-2)_\mu/2$.
The measurement was by the muon $g-2$ collaboration at Brookhaven National Laboratory~\cite{g-2_bnl2010}, and the Standard Model prediction was obtained from the combination of the results in Refs.~\cite{Aoyama:2012wk,g-2_EWderafael2002,g-2_EWmarciano2002,g-2_Hagiwara2011,Prades:2009tw} (also see Refs.~\cite{g-2_davier2010}).
This discrepancy can be interpreted as another signal of physics beyond the Standard Model.

The SUSY is capable to solve the $\gmt$ anomaly~\cite{SUSY_gminus2}.
The MSSM dominant contribution to $\gmt$ is, as summarized in FIG.~\ref{fig:MSSM_g-2}, one-loop level diagrams with smuon--neutralino ($\tilde \mu$--$\tilde\chi^0$) and muon-sneutrino--chargino ($\tilde\nu_\mu$--$\tilde\chi^-$).
They are respectively expressed as~\footnote{We take the convention where the SUSY-breaking gaugino masses are taken to be positive: $M_a>0$.}
\begin{align}
 \Delta a_\mu(\tilde\mu,\tilde\chi^0)&\approx \frac{\alpha_Y m_\mu^2}{m^2\s{soft}}\sgn(\mu)\tan\beta+\cdots,
&
 \Delta a_\mu(\tilde\nu_\mu,\tilde\chi^\pm)&\approx \frac{\alpha_2 m_\mu^2}{m^2\s{soft}}\sgn(\mu)\tan\beta,
\end{align}
where $\alpha_Y$ and $\alpha_2$ are the gauge coupling strengths of $\rm{U}(1)_Y$ and $\rm{SU}(2)\s{weak}$, $\mu$ is the $\mu$-term (the Higgsino mass term), and $\tan\beta=\vev{H\s{u}}/\vev{H\s{d}}$.
$m\s{soft}$ represents the mass of the loop-going superparticles.
In order to suffice $\Delta a_\mu$ with these two diagrams, the corresponding superparticles should be (precisely, at least ($\tilde\mu_1$,$\tilde\chi^0_1$)-pair, or ($\tilde\nu_\mu$,$\tilde\chi^\pm_1$)-pair should be) as light as $\Order(100)\GeV$, and $\tan\beta=\Order(10)$ is favored.

Here a naive idea comes up: the squarks are as heavy as $\Order(1\text{--}10)\TeV$ to raise the Higgs mass to $126\GeV$ and to escape from the LHC SUSY search, meanwhile some of the $\gmt$-related superparticles are very light as $\Order(100)\GeV$.
We will call this scenario ``$\gmt$-motivated MSSM'', and investigate it in this article.

The discussion performed here will be restricted to phenomenological approach.
It is interesting and of great importance to construct viable SUSY models which realize such mass separation, but we will put it out of the scope.
We treat the masses of the superparticles as free parameters and discuss the following questions with model-independent approach:
\begin{enumerate}
 \item Is this scenario still viable? --- The answer is YES; i.e., the SUSY can still explain the $\gmt$ anomaly (with optimism that suitable models for this scenario will be invented~\footnote{For an example, see Ref.~\cite{Ibe:2012qu}.}).
 \item Can we investigate this scenario at the LHC and future colliders? --- We will see that most of the parameter space which can explain the $\gmt$ anomaly can be covered with the LHC, but some regions are difficult to be searched for, for which we should develop the ways to search.
\end{enumerate}

The rest of this article is composed as follows: Section~\ref{sec:2} is for the declaration of the $\gmt$-motivated MSSM, Section~\ref{sec:3} is the discussion on LHC phenomenology, and they are summarized in Section~\ref{sec:4}.

\section{$\boldsymbol{\gmt}$-motivated MSSM}\label{sec:2}
Here we clarify the targeted model with utilizing several simplifications.
The squarks are set much heavier than $1\TeV$ as required by the Higgs mass.
The slepton soft masses are assumed to be diagonal and common for the first and the second generations: $(m^2_L)_1=(m^2_L)_2$ and $(m^2_{\bar E})_1=(m^2_{\bar E})_2$, while the third generation sleptons are set decoupled to simplify the collider phenomenology: $(m^2_L)_3\sim(m^2_{\bar E})_3=\Order(1)\TeV$.
For the gaugino masses $M_a$, an approximate GUT relation $M_1:M_2:M_3=1:2:6$ is utilized.
The scalar trilinear couplings ($A$-terms) are simply set zero.
The Higgs sector is set as $(m_A,\tan\beta)=(1.5\TeV,40)$ for a large $\tan\beta$ is preferred.
The (lighter $CP$-even) Higgs mass is just set by hand as $m_h=126\GeV$, which is assumed to be realized by the heavy squark masses.

Here it should be emphasized that most of the above simplifications/assumptions do not affect LHC phenomenology, or do ease the LHC SUSY searches. The exceptions are the gaugino mass relation and the decoupling tau-slepton sector.
Especially, LHC phenomenology with lighter tau-sleptons is left as future works.

Finally, $(m^2_L,m^2_{\bar E},M_2,\mu)$ are left as free parameters.
In the original paper \cite{Endo:2013bba}, we performed analyses on the four parameter spaces:
\begin{itemize}\itemindent20px\labelsep10px
 \item[(a)--(c)] $\left(M_2,\sqrt{m^2_{L}}\right)$-plane, where $m_{\bar E}^2$ is set as $(3\TeV)^2$, and $\mu$ is as (a)$M_2$, (b)$2\times M_2$, and (c)$0.5\times M_2$.
 is set as $(M_2,3\TeV)$, $(2M_2,3\TeV)$, $(0.5M_2,3\TeV)$, respectively.
In these cases, since the right-handed sleptons are decoupled, diagrams with Higgsino becomes much smaller, and the SUSY contribution to $\gmt$ is dominated by the chargino--muon-sneutrino diagram (FIG.~\ref{fig:MSSM_g-2}-right).
 \item[(d)]
$\left(M_2,\sqrt{m^2_{L}}\right)$-plane, where $m_L^2 : m_{\bar E}^2=2^2:3^2$, and $\mu=2\TeV$.
This case is somewhat special for the diagram with a bino--smuon loop dominates the SUSY contribution.
This is because the large $\mu$-term enhances the $\mu\s L$--$\mu\s R$ mixing and suppresses the diagrams with Higgsinos.
As a result, the $\gmt$ anomaly can be explained even if bino is as heavy as $500\GeV$.
\end{itemize}
In this article, however, we just consider the model (d), the most interesting one, for simplicity.

\section{$\boldsymbol{\gmt}$-motivated MSSM v.s.~LHC}\label{sec:3}
\subsubsection{Overview}
The LHC SUSY search for this scenario is summarized to two schemes.
The first one targets gluinos, hopefully which sit around $1\TeV$ for the approximate GUT relation.
The gluinos can be searched for with the well-known strategy, i.e., searches for events with multiple hard jets and a large missing energy.
An example is in Ref.~\cite{ATLAS2012109}, where the ATLAS collaboration analyzes their data corresponding to an integrated luminosity of $5.8\invfb$ obtained at $\sqrt{s}=8\TeV$ to constrain the gluino mass as $m_{\tilde g}>950\GeV$ when squarks are decoupled.
However, these searches are hopeless with gluino decoupled, i.e., without the GUT relation, which was not necessary for the SUSY explanation of the $\gmt$ anomaly.

The other scheme focuses on direct productions of the $\gmt$-related superparticles.
It faces more Standard Model background events compared with the first scheme, but is very important since with such searches we can distinguish whether the SUSY is still viable as the solution to the $\gmt$ anomaly or not.
Especially, the multi-lepton signatures are particularly important, because they are provided by the $\gmt$-related superparticles.
As a result which focuses on direct productions of electroweakinos (charginos and neutralinos), the ATLAS collaboration recently reported searches for events with three leptons plus a large missing energy in the data of $13.0\invfb$ at $\sqrt{s}=8\TeV$~\cite{ATLAS2012154,999}\endnotetext[999]{%
One month after this talk, the ATLAS collaboration updated this report (Ref.~\cite{ATLAS2012154}; $8\TeV$, $13.0\invfb$ data) with a report in which they analyzed the data corresponds to $20.7\invfb$ obtained at $\sqrt{s}=8\TeV$~\cite{ATLAS2013035}.
They observed no excess above the standard model expectations, and provided tighter exclusion limits on models beyond the Standard Model than they did with $13.0\invfb$ data.
In this article, however, this result is not considered at all.}, where multi-leptons come from electroweakinos and a large missing energy from the lightest superparticle (LSP).

In this article, we interpret these two results from the ATLAS collaboration, i.e., the multi-jet search in Ref.\cite{ATLAS2012109} targeting gluino pair production, and the multi-lepton search in Ref.\cite{ATLAS2012154} focusing on the electroweakino pair production, to obtain constraints on the $\gmt$-motivated MSSM.

\subsubsection{Method}
We performed Monte Carlo simulation to interpret the ATLAS results.
For each model point, mass spectrum is generated with {\tt SOFTSUSY\,3.4}~\cite{SOFTSUSY} and {\tt SUSY-HIT\,1.3}~\cite{SUSYHIT}, and the SUSY contribution to the muon $g-2$ is calculated with {\tt FeynHiggs}~\cite{FeynHiggs},
SUSY events are generated by {\tt Pythia\,6.4}~\cite{Pythia6.4} with the {\tt CTEQ6L1} set of parton distribution functions (PDFs), and normalized with the next leading order cross section; for gluino pair production it was obtained {\tt Prospino\,2}~\cite{Prospinoweb,ProspinoSG}, where the {\tt CTEQ6L1} and {\tt CTEQ6.6M} PDFs~\cite{PDFCTEQ6} are used, and for the electroweak channels the factor is set as $K=1.2$, which is a typical value in the parameter space.
{\tt Delphes\,2.0}~\cite{Delphes} was used to simulate detector response.

Efficiency of triggering is not considered, but efficiency and fake rate of $b$-tagging, and efficiency of lepton detection was taken into account.
Detailed description of our efficiency estimations as well as object definitions can be found in Ref.~\cite{Endo:2013bba}.

The signal regions are defined to be the same as those in the original ATLAS analyses.
The $CL\s{s}$ method is used to derive exclusions for each model point.
The numbers of simulated SUSY events in the signal regions are compared to the corresponding upper bounds obtained in the ATLAS reports.
The analysis procedures are validated by comparing the simulations with the ATLAS results.

\subsubsection{Result}
The result for case (d) is shown in FIG.~\ref{fig:result}-left.
It should first be emphasized that the region where the $\gmt$ anomaly, drawn with an yellow/orange band, extends transversely to the right edge of the plotted region, which is the special feature of the case (d) originating from the enhancement of the bino--smuon diagram due to the large left-right mixing of smuons.
The $\gmt$ anomaly is explained even at $M_2=1.3\TeV$ for $m^2_L=(200\GeV)^2$ at the $1\sigma$ level.
As we will discuss below, the case (d) is difficult to be covered fully at the LHC for this feature.

The gray region is already excluded by the LHC SUSY searches (see Note~\cite{999}) at 95\% confidence level.
The leftmost region, where the gluino {\em pole} mass is lighter than $\sim950\GeV$, is excluded by the multi-jet search as expected.
The center of the plotted region is covered by the multi-lepton search.
The shape of the region seems somewhat strange, but is understood easily; it is because the ATLAS multi-lepton search focuses the region in which $\tilde\chi^0_1<m_{\tilde l^\pm}<\tilde\chi^0_2$, where the direct pair production such as $pp\to\tilde\chi^\pm_1\tilde\chi^0_2$ can provide three leptons as, e.g.,
\begin{align}
 \tilde\chi^\pm_1&\to\tilde\nu+l^\pm\to\tilde\chi^0_1+l^\pm+\nu,
&
 \tilde\chi^0_2&\to\tilde l^\mp+l^\pm\to\tilde\chi^0_1+l^\pm+l^\mp.
\end{align}

In FIG.~\ref{fig:result}-left, the gluinos are assumed to be within the reach, but they can be decoupled from the viewpoint of the $\gmt$ anomaly.
The result with gluinos decoupled is provided as FIG.~\ref{fig:result}-right; in the figure, no regions are excluded by the multi-jet search, for hard jets are less expected without gluinos. Instead, the leftmost region is excluded by the multi-lepton search.
The excluded region in the center of the parameter space is the same as FIG.~\ref{fig:result}-left as expected.

\subsubsection{Discussion / Prospects}
Now we realized that the SUSY is still viable as the solution to the $\gmt$ anomaly.
Then, how can we cover the remaining regions?
Here, focusing on this question, let us briefly discuss prospects.
(Detailed discussion can be found in Ref.~\cite{Endo:2013bba}.)

\begin{figure}[t]
  \begin{center}
     \includegraphics[width=0.9\textwidth]{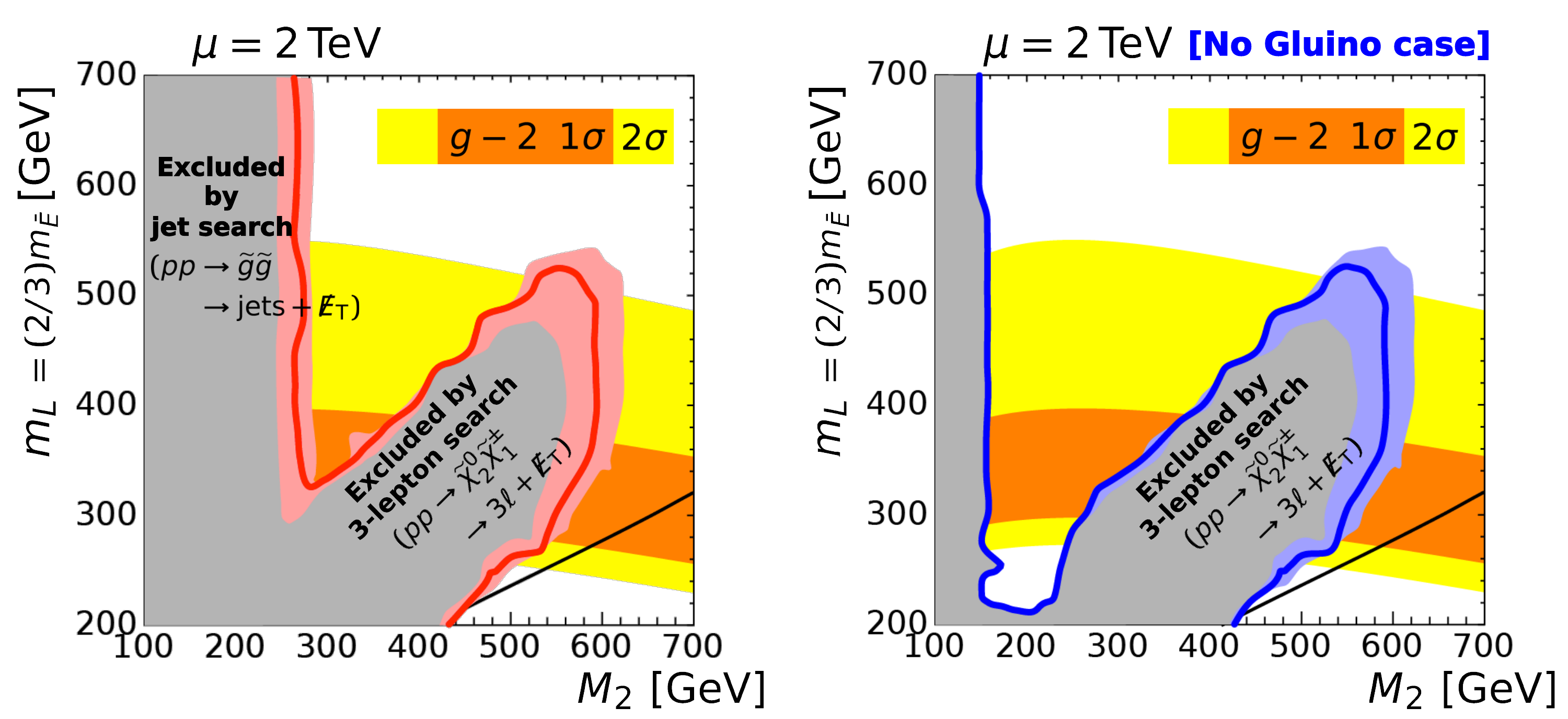}\\(d) $\mu = 2\TeV $, $m^2_{L}:m^2_{\bar E} = 4:9$
  \end{center}
  \caption{
Current LHC bounds on the $\gmt$-motivated MSSM.
Here the result of the case (d) is shown; those of (a)--(c) can be found in Ref.~\cite{Endo:2013bba}.
In the left figure, the gluinos are as light as expected from the approximate GUT relation, while they are assumed to be decoupled in the right.
The orange (yellow) bands show the region where the $\gmt$ anomaly is explained by the SUSY contributions at the $1\sigma$ ($2\sigma$) level.
The red and blue lines show the current bounds from the LHC SUSY searches~\cite{ATLAS2012109,ATLAS2012154} (see Note~\cite{999}); the gray regions are excluded at 95\% confidence level, and the theoretical uncertainty of $\pm30\%$ is depicted by the red and blue bands.
The LSP is $\tilde\chi^0_1$ ($\tilde\nu$) in the regions above (below) the black thick lines.
}
  \label{fig:result}
\end{figure}

If the approximate GUT relation on the gaugino masses realizes (Fig.~\ref{fig:result}-left), SUSY searches focusing on multi-jet events at the 13--14\,TeV LHC can cover the gap between the excluded regions.
However, the gap cannot easily be covered without gluinos (Fig.~\ref{fig:result}-right).
In the gap region, the produced electroweakinos decay into $W/Z$ bosons so that the electroweakino pair production results in similar events as the Standard Model di-boson production.
(Actually, the leftmost region in the right figure is excluded because the electroweakinos decay into three bodies.)
We should develop the method to distinguish such events from the Standard Model background events with utilizing the large missing energy, or rely on slepton pair production even though the cross section is smaller than electroweakino pair production.

The rightmost regions of the figures are much more challenging.
Although slepton pair production is available with a considerable rate at the LHC, it just results in di-lepton signature and the discrimination from the Standard Model events is extremely difficult.
The $M_{T2}$ method~\cite{MT2} might be, from a naive thought, useful for the discrimination, but the ILC is more suitable to explore this parameter region.

\section{Summary}\label{sec:4}
The SUSY is the most promising candidate for physics beyond the Standard Model, since it solves the hierarchy problems, provides a dark matter candidate, realizes the Higgs mass of $126\GeV$, and explain the $\gmt$ anomaly.
However, the LHC SUSY search as well as the $126\GeV$ Higgs boson indicates that the superparticles, especially squarks and gluinos, are not so light as we expected.
Now it is important to investigate the SUSY searches which are not involved with the colored superparticles.

As a clue for the investigation, we focused on the $\gmt$ anomaly, which can be explained by the SUSY contributions if some of neutralinos, charginos, and muon-sleptons are as light as $\Order(100)\GeV$.
We proposed the $\gmt$-motivated MSSM as a benchmark model, and interpreted the up-to-date results of the LHC SUSY search~\cite{ATLAS2012109,ATLAS2012154} to obtain the experimental bound on the model.
In this article, we discussed only the case (d), the most challenging benchmark model.

The results are shown in FIG.~\ref{fig:result}; the left figure is under the assumption that $M_1:M_2:M_3=1:2:6$, where the SUSY search focusing on the gluino pair production \cite{ATLAS2012109} excludes the leftmost region, and the right figure is with gluino decoupled.
We saw that the multi-lepton search~\cite{ATLAS2012154} is capable to exclude a wide region regardless of presence of the gluino.
However, at the same time, we realized that two sorts of the mass spectra remain viable.

These two regions can be regarded as the targeted regions in the future colliders.
The difficulty of the gap region comes from the feature that the SUSY events are similar to the Standard Model di-boson production, while the rightmost region is extremely challenging since only the di-lepton signatures from slepton direct pair production are expected as the SUSY signature.

Finally let us mention what we did not cover/discuss in this article.
First, we assume in order to simplify LHC phenomenology that staus and tau-sneutrinos are decoupled; analyses and discussion with lighter tau-sleptons are of interest for it is more natural, and left as future works.
Also the fixed ratio of the gaugino masses, $M_1:M_2=1:2$, can be relaxed.
This is important for the study on the rightmost region of FIG.~\ref{fig:result}, since $M_2$ is irrelevant for $\gmt$ in the case (d) but LHC phenomenology in the region is considerably affected by the mass splittings between sleptons and electroweakinos.


\bigskip
\begin{acknowledgments}
 Author is grateful to the collaborators of this project~\cite{Endo:2013bba}, Dr.~Endo, Prof.~Hamaguchi and Mr.~Yoshinaga.
  He would like to thank the organizers of the conference HPNP\,2013 at University of Toyama for their warm hospitality and giving him the opportunity to have a talk.

This work was supported by JSPS KAKENHI Grant No.~22--8132.
\end{acknowledgments}

\bigskip

\def\href#1#2{#2}

\bibliography{ExpResLHC,MagneticMoment,HEPComputing,SUSYHiggs,ExpLHC,bmeson,vectorlike,SUSY,LHCPhenom,DarkMatter}

\end{document}